\documentstyle[psfig,aps,prl,twocolumn]{revtex}
\begin{document}
\draft \author{Frank Sander, Thibaut Devolder, Tilman Esslinger, and
Theodor W.~H\"ansch} 
\address{Sektion Physik,
Ludwig-Maximilians-Universit\"at, Schellingstr.\ 4/III, D-80799 Munich,
Germany and\\ Max-Planck-Institut f\"ur Quantenoptik, D-85748
Garching, Germany} 
\date{\today} 
\title{Ramsey type Sub-Recoil Cooling}
\maketitle
\begin{abstract}

\noindent 
We experimentally study the motion of atoms interacting with a
periodically pulsed near resonant standing wave. For discrete pulse
frequencies we observe a comb-like momentum distribution. The peaks
have widths of $\approx 0.3\,\hbar k$ and a spacing which is an
integer multiple of the recoil momentum $\hbar k$. The atomic
population is trapped in ground states which periodically evolve to
dark states each time the standing wave is switched on.

\end{abstract}


\noindent 
Ultra cold atoms form an ideal system to study the evolution of matter
waves. The cold atoms have a large spatial coherence and can be
manipulated by atom-light interaction. Using laser light, motional
quantum states of atoms can be precisely prepared and observed. Very
recently, Bloch oscillations \cite{BlochOscillations} and the
Wannier-Stark ladder \cite{Raizen1} have been demonstrated in
fascinating experiments, where laser cooled atoms interacted with an
accelerating standing wave potential. A similar system has been used
to study the relation between quantum evolution and the underlying
classical dynamics \cite{Raizen2}. These experiments were performed in
a regime where dissipation due to spontaneous emission was negligible.

In this work we study the evolution of atomic matter waves interacting
with a periodically pulsed standing wave in the presence of
dissipation.  We prepare atoms via spontaneous emission in
non-stationary quantum states, which are superpositions of different
momentum eigenstates.  We monitor their evolution and observe revivals
of states which are decoupled from the light field by varying the time
between successive pulses. The periodic interaction with the standing
wave leads to an accumulation of the atoms in a comb-like momentum
distribution with peaks narrower than the recoil momentum of a single
photon.  Our scheme can thus be regarded as a new approach to
sub-recoil cooling. In velocity selective coherent population trapping
(VSCPT) \cite{VSCPTExp} and Raman cooling \cite{RamanCooling} the
velocity selection is achieved during the interaction with either a
continuous light field or with a velocity-selective Raman pulse. For
both techniques a long interaction time with the light field results
in narrow momentum distributions. In our scheme the kinetic
evolution of the atoms between two successive pulses leads to the
selection of sharply defined velocity classes. This is similar to
Ramsey spectroscopy \cite{Ramsey}, where the energy difference between
atomic states is determined using two or more separated interaction
regions instead of a single large one.

To understand the interaction of the atoms with the pulsed standing
wave let us recall the concept of dark states. Consider an atomic
transition which has an integer total angular momentum $F$ both in the
ground and excited state manifolds. If this transition is driven with
polarized light, there is one ground state of the $2F+1$ ground states
which is not coupled to the excited state manifold \cite{Russen}. This
state is called a dark state and it can be populated by optical pumping. 

If the atom interacts with a continuous one dimensional standing wave
field with a polarization gradient a slightly more complex situation
arises.  We then have to include the motion of the atom along the
standing wave axis. For the special case of an atom with a $F=1 \to
1$ transition there is a dark state which is also an eigenstate of
the kinetic energy \cite{Maxim}. This stationary dark state
facilitates VSCPT cooling. For atoms with larger angular momenta
stationary dark states are not found in the considered light field,
{\sl i.~e.} there are no dark states which are also eigenstates of the
kinetic energy. Consequently the atom cannot remain decoupled from the
light field \cite{Arimondo1}.

In the case of the periodically pulsed standing wave the atom
kinetically evolves in the time between successive pulses. The kinetic
evolution can lead to a revival of the dark
state\cite{Arimondo2,Weitz}, where the revival time depends on the
atomic velocities along the standing wave axis. This revival of the
dark state can be periodic. An atom can therefore propagate through
the pulsed standing wave without being excited if it is in a state
that evolves to a dark state each time the standing wave is switched
on. These states are superpositions of states with different momenta
and will be referred to as propagating dark states. They exist only
for discrete pulse frequencies (for $F>1$) and have discrete
momenta. The propagating dark states are populated by optical pumping
and we expect the atomic population to accumulate in these states
after a sufficiently long sequence of pulses. The atomic momentum
distribution will then show the momenta of propagating dark states,
which results in a comb-like distribution.

\begin{figure}
\centerline{\psfig{file=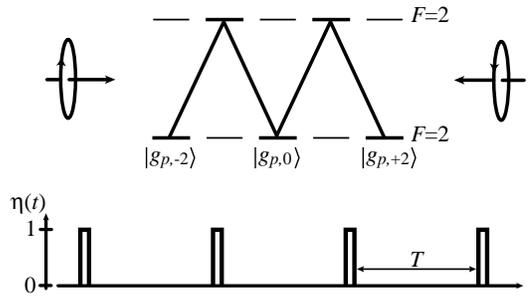,angle=0,width=0.39\textwidth}}
\caption {{Coherent coupling within a momentum family. $|g_{{\tiny p,
m}}\rangle$: ground state with the magnetic quantum number $m$ and the
momentum $p+m \hbar k$. Below: time dependence of the atom-light
interaction in the pulsed standing wave.}}
\label{Levelscheme}
\end{figure}

To be more specific consider a $F=2\to2$ transition interacting with a
periodically pulsed standing wave. The total Hamiltonian of this
system is given by
$$H=P^2/(2M)+H_{A}+\eta(t)V,$$ where $P$ is the atomic momentum
operator, $M$ the atomic mass and $H_A$ the Hamiltonian of the
internal atomic states.  The function $\eta(t)$ is plotted in the
lower part of Fig.~\ref{Levelscheme} and describes the time
dependence of the atom-light interaction $V$. It is periodically
switched on for time intervals of the length $\tau$ separated by
longer intervals of the length $T$. The standing wave consists of two
counter propagating waves with circular polarizations inducing
$\sigma^+$ and $\sigma^-$ transitions. The atom-light interaction can
be written as a sum over different momentum families \cite{VSCPTTheo}:
\begin{eqnarray*}
V&=&\frac{1}{2 }\hbar \Omega e^{-i\omega_{\tiny L} t}\sum_p V_{\tiny
p}/{\tiny \sqrt{6}} +\mbox{h.c.} ,
\end{eqnarray*}
where $\Omega$ is the resonant Rabi coupling and $\omega_{\tiny L}$
the laser frequency. The term $V_{\tiny p}$ describes the coherent
coupling within a momentum family. It consists of a W-type and a
M-type coupling scheme, as illustrated in the upper part of Fig.~1. We
will neglect the W-type system (thin lines) since it does not
contribute to the dark state and is depopulated by optical
pumping. The M-type coupling scheme (bold lines) is given by
\begin{eqnarray*}
V_{\tiny p}&=&\sqrt{2}|e_{\tiny p,-1}\rangle \langle g_{\tiny
p,-2}|-\sqrt{3}|e_{\tiny p,-1}\rangle \langle g_{\tiny p,0}|\\
&&+\sqrt{3}|e_{\tiny p,+1}\rangle \langle g_{\tiny
p,0}|-\sqrt{2}|e_{\tiny p,+1}\rangle \langle g_{\tiny p,+2}| ,
\end{eqnarray*}
where $|g_{{\tiny p, m}}\rangle$ ($m=\pm2, 0$) describes a
ground state with the magnetic quantum number $m$ and the momentum
$p+m \hbar k$ (where $k$ is the wave vector). Correspondingly,
$|e_{{\tiny p, m}}\rangle$ ($m=\pm1$) describes an excited state with
the magnetic quantum number $m$ and the momentum $p+m \hbar k$. The
magnetic quantum numbers label the eigenstates of the projection of
the total angular momentum $\vec{F}$ on the axis of the standing
wave. Different momentum families are coupled only via spontaneous
emission.

Let us now consider the ground state $\psi_{\tiny p}$ 
\begin {eqnarray*}
\left|\psi_{\tiny p}\right\rangle 
&=& e^{-i\phi_{{\tiny p, -2}}} \sqrt{\scriptstyle
\frac{3}{8}}|g_{\tiny p, -2}\rangle \\
&&+ e^{-i\phi_{{\tiny p, 0}}} \sqrt{\scriptstyle
\frac{1}{4}}|g_{\tiny p,  0}\rangle\\
&&+ e^{-i\phi_{{\tiny p, +2}}} \sqrt{\scriptstyle
\frac{3}{8}}|g_{\tiny p,  +2}\rangle.
\end {eqnarray*}

\noindent It is a superposition of three states with different momenta
and different magnetic quantum numbers, which belong to the same
momentum family $p$. If the phases $\phi_{\tiny p, 0}$ and
$\phi_{\tiny p, \pm 2}$ have the same value (modulo $2\pi$), the state
$\psi_{\tiny p}$ is a dark state and does not couple to the light
field, {\sl i.~e.} $V|\psi_{\tiny p}\rangle=0$. It is not an eigenstate of
the kinetic energy and therefore not a stationary dark state. The free
kinetic evolution of $\psi_{\tiny p}$ can be described by the time
dependent phases $\phi_{\tiny p, 0}(t)$ and $\phi_{\tiny p, \pm 2}(t)$
which evolve according to the kinetic energy of the corresponding
momentum states. These phases are given by $\phi_{\tiny p, 0}(t)
=\left(p/(\hbar k)\right)^2\omega_rt$ and $\phi_{\tiny p, \pm
2}(t)=\phi_{\tiny p, 0}(t)+ 4\omega_rt\pm 4\omega_rp/(\hbar k)t$,
where $\omega_r=\hbar k^2/(2M)$ is the recoil frequency. Here we
assume that the state $\psi_{\tiny p}(t)$ is in a dark state at $t$=0
with $\phi_{\tiny p, 0}(t$=0$)=\phi_{\tiny p, \pm 2}(t$=0$)=0 $. After
a revival time $T$ the state $\psi_{\tiny p}$ is again a dark state if
the phases satisfy $\phi_{\tiny p, 0}(T)-\phi_{\tiny p, -2}(T)=2\pi m$
and $\phi_{\tiny p, +2}(T)-\phi_{\tiny p, 0}(T)=2\pi (m+n)$, where $m$
and $n$ are integer numbers. Both conditions are satisfied for
times $T_n=n\tau_r/8$ (where $\tau_r=2\pi/\omega_r$ is the recoil
time), if $\psi_{\tiny p}$ belongs to a momentum family with
$p=p^{(n)}_m=\frac{m}{n}2\hbar k+\hbar k$. The kinetic evolution of a
state $\psi_{\tiny p}$ (with $p=p^{(n)}_m$) is periodic and it is a
dark state at times $t=0, T_n, 2T_n, 3T_n,\dots$. If the standing wave is
switched on only at these times ({\sl i.~e.} when $\psi_{\tiny p}$ is a dark
state), $\psi_{\tiny p}$ is a propagating dark state and its
interaction with the standing wave field vanishes. For $n=1$ the
propagating dark states have momenta which are odd integer multiples
of $\hbar k$. For $n=2$ the propagating dark states have momenta which
are even and odd integer multiples of $\hbar k$ \cite{comment}. For
$n>2$ propagating dark states can also have momenta which are integer
fractions of $\hbar k$ \cite{OpticalDiffraction}.

The lifetime of a propagating dark state in the periodically pulsed
standing wave is limited by the intervals $\tau$ during which the
standing wave is switched on. During this time the kinetic Hamiltonian
leads to an effective coupling of the dark state to the light field
\cite{VSCPTTheo,Arimondo1}. This results in a lifetime $1/\Gamma''$ of
the dark state which is proportional to
$\frac{|\Omega|^2}{\Gamma}(1+\frac{2\Delta}{\Gamma})^{-1} $ (where
$\Gamma^{-1}$ is the excited state life time and $\Delta$ the detuning
between laser and atomic transition frequency). To minimize the
excitation of a propagating dark state it is necessary to choose
$\tau \ll 1/\Gamma''$. 
A high Rabi coupling effectively hinders the evolution
of the propagating dark state \cite{why}. We therefore expect the
longest lifetime for propagating dark states when the time between
successive standing wave pulses is $T=T_n$ (rather then $\tau
+T=T_n$). This corresponds to the assumption that we can neglect the
evolution of the propagating dark state while the standing wave is
switched on.

Atoms which are not in propagating dark states are excited by the
pulsed standing wave. Each spontaneous emission of a photon changes
the momentum of the atom, and the atom can decay to a propagating dark
state. There it experiences a strongly reduced excitation rate. We
therefore expect that atoms accumulate in propagating dark states
during the interaction with the pulsed standing wave. The momentum
distribution will then show peaks at the momenta of propagating dark
states. The width of the peaks can become narrower than the single
photon recoil. This velocity selectivity is due to the kinetic
evolution of atoms between two pulses. Consider an atom in a state
$\psi_{\tiny p'}$ which is in a dark state at $t$=0 and has a center
momentum $p'=p^{(n)}_m + \delta p$ ($\delta p \ll \frac{m}{n}2\hbar k$). During the
free kinetic evolution the state $\psi_{\tiny p'}$ will increasingly
deviate from the propagating dark state $\psi_{\tiny p}$ (with
$p=p^{(n)}_m$). When the standing wave is switched on again after a
time $T=T_n$, the state $\psi_{\tiny p'}$ has not completely evolved
to a dark state. This increases the probability that $\psi_{\tiny p'}$ is
excited by the standing wave and it reduces the atomic population in
the momentum family $p'$.

\begin{figure}
\centerline{\psfig{file=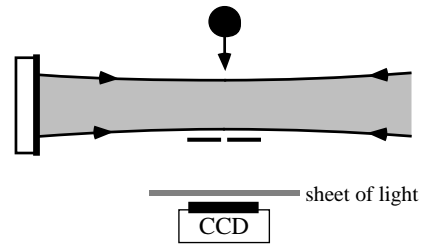,angle=0,width=0.30\textwidth}}
\caption {Experimental set-up}
\label{ExperimentalSetup}
\end{figure}

To investigate the dynamics of propagating dark states, we have
performed an experiment with $^{87}$Rb atoms (Fig.\
\ref{ExperimentalSetup}). A cloud of ${\tiny \approx}10^8$
magneto-optically trapped atoms is accelerated downwards using optical
molasses cooling to a moving reference frame. After 17\,cm of flight
the cloud arrives with a speed of 3.2\,m/s and a density of $n\approx
10^8\,{\rm cm}^{-3}$ in the interaction region, which is
shielded with mu-metal against magnetic fields to below 0.5\,mG. On
their ballistic trajectory downwards the atoms interact with a
horizontally aligned $\sigma^+-\sigma^-$ standing wave having a
vertical Gaussian waist of 1.37\,mm. The light field is tuned
$\Delta=14\,\Gamma=2\pi\cdot 40\,$MHz to the blue of the $F=2\to 2$
transition of the D$_1$ line, where $\Gamma^{-1}=28\,$ns is the life
time of the excited states. Each running wave has a peak intensity of
9.4\,mW/cm$^2$, which corresponds to a resonant Rabi coupling of
$\Omega=1.3\,\Gamma$ and to an excitation rate of $\Gamma'=8\cdot
10^{-3}\,\Gamma=(4\mu{\rm s})^{-1}$ (on the $F=2\to 2$ transition).

An accusto-optical modulator is used to switch the standing wave on
for intervals of $\tau=\tau_r/100=3\,\mu$s alternating with dark
intervals of $T=\tau_r/8=35\,\mu$s (corresponding to $n=1$). During
the interaction time of 1.1 ms the atoms are subjected to 28 standing
wave pulses. To recycle atoms that have decayed to the $F=1$ ground state
manifold a continuous standing wave of 1.6\,mm waist overlaps. It is
tuned to the $F=1\to 2$ transition of the D$_2$ line. Its single pass
Rabi coupling is $\Omega_{12}=0.2\,\Gamma$, which corresponds to an
excitation rate of $0.02\,\Gamma=(1\,\mu{\rm s})^{-1}$. The
interaction time with this additional light field is 1.5\,ms. Both
light fields are derived from grating stabilized laser diodes. The
laser beams are spatially filtered to achieve Gaussian modes. The
standing waves are formed by retroreflection off a mirror, which is
13\,cm away from the interaction region. A quarter wave plate in front
of the mirror is used to provide the $\sigma_+-\sigma_-$-polarization
of the standing wave.

To determine the momentum distribution we place a pinhole of
$75\,\mu$m diameter 5\,mm below the standing wave axis. The atoms that
pass through the pinhole ($\approx\!3\cdot 10^3$ atoms) expand
horizontally in two dimensions according to their transversal
momentum. A transversal momentum of $1\,\hbar k$ translates to a
$170\,\mu$m transversal displacement in a plane 9.6\,cm below the
pinhole.  The spatial distribution of the atoms in this plane is
imaged by recording the fluorescence in a sheet of light with a CCD
camera. The sheet of light is formed by a standing wave, which is
resonant with the closed $F=2\to 3$ transition of the D$_2$ line. This
allows detection of atoms in the $F=2$ ground state. Optionally, an
additional laser beam 2\,mm above the sheet of light optically pumps
the atoms from the $F=1$ to the $F=2$ ground state manifold and can be
used to additionally detect atoms in the $F=1$ ground state. We have
used this beam to verify that no atoms leave the interaction region in the
$F=1$ ground state manifold. The overall momentum resolution of the
detection system has been improved to $\sigma=0.3\,\hbar k$ (where
$\sigma$ is the $e^{-1/2}$ half width), as compared to previous
experiments done with the same apparatus \cite{VSCPTTransfer}.  The
measurements presented here are integrated over 200 atom clouds
extracted from the magneto-optical trap at a rate of $1\,{\rm
s}^{-1}$. The stray light background has been measured in an identical
repetition of the experiment but without atoms and has been subtracted
from the data.  To yield one dimensional momentum distributions we
have integrated the two dimensional images along the direction
perpendicular to the cooling axis.

\begin{figure}
\centerline{\psfig{file=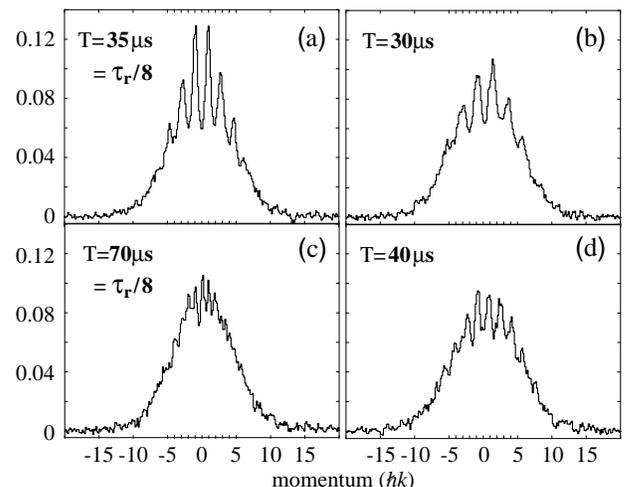,angle=0,width=0.45\textwidth}}
\caption {Momentum distributions obtained for different separations of the standing wave pulses. See text. Vertical axis: momentum
space density in units of $(\hbar k)^{-1}$.}
\label{Plot}
\end{figure}

In Fig.~\ref{Plot}(a) the momentum distribution measured for a pulse
spacing $T=35\mu$s is shown. The comb-like structure with $2\hbar k$
spacing between the sub-recoil cold peaks stems from atoms trapped in
propagating dark states. The peaks occur at odd multiples of $\hbar
k$, as we expect for n=1 ($T=1\tau_r/8=35\,\mu$s). Each propagating
dark state contributes to three neighboring peaks. The width of these
peaks is determined by a best Gaussian fit as $\sigma=0.3\,\hbar k$.
The envelope of the momentum distribution has a width of
$\sigma\approx 4.4\,\hbar k$. It results from broadening of the
initial distribution, which has for geometrical reasons a width of
$\sigma=1.7\,\hbar k$ (measured with all fields permanently switched
off in the interaction region).  We experimentally varied the length
$\tau$ of the light pulses. For pulses up to a factor of two longer
the observed momentum distribution did not change significantly,
whereas shorter pulses lead to a reduced contrast in the momentum
spectrum. This is in agreement with the calculated $\tau\Gamma'=0.8$
excitation cycles per light interval.  Fig.~\ref{Plot}(b) and
Fig.~\ref{Plot}(d) show the momentum distributions for pulse spacings
$T=30\,\mu$s and $T=40\,\mu$s. The sharp peaks smear out since in both
cases the pulse spacing deviates from the resonance condition
$T=35\,\mu$s. Propagating dark states which are completely decoupled
from the standing wave do not exist for these pulse spacings. A
further increase of the pulse spacing to $T=60\,\mu$s completely
washes out the comb-like structure. For $T=2\tau_r/8=70\,\mu$s
[see Fig.~\ref{Plot}(c)] a momentum comb appears again with a spacing of
$\,\hbar k$, as expected for the $n=2$ resonance. 

Our scheme can be extended to two and three dimensions and to
transition schemes of the type $F\to F$ and $F\to F-1$ (for $F>1$). For
$\Lambda$-type coupling schemes the non-coupling state is a superposition
of two momentum states, so that for any time between successive pulses
a propagating dark state exists. As a cooling technique our scheme
might find particular interest for atoms which have transition
frequencies that can only be excited using pulsed laser-sources,
e.g. the Lyman-$\alpha$ transition of hydrogen. As our scheme provides
continuous cooling it can be applied to atomic beams so that another
tempting application appears in atom lithography. There the width of
the fabricated nanostructures is mainly limited by the transversal
momentum distribution of the atomic beam.

In conclusion, we have experimentally shown that propagating dark
states can be populated if the standing wave is pulsed with a
frequency of $1\tau_r/8$ or $2\tau_r/8$. For larger time intervals
($n\tau_r/8, n\ge 3$) between the pulses we expect the atoms to
accumulate in momentum states which are spaced by fractional multiples
of $\hbar k$.

\acknowledgements 

\noindent We wish to thank A. Lambrecht and H. Ritsch for
helpful discussions and the Deutsche Forschungsgemeinschaft for
support.

\begin{references}
\bibitem{comment} The evolution of atoms precooled by VSCPT which are
subjected to a delayed pulse is theoretically studied in
Ref. \protect\cite{Arimondo2}. Their simulation of the $F=2$ system
corresponds to the special case of $n=4$ and $m=-2$.
\bibitem{why} This is analogous to the situation in VSCPT cooling, 
where a high Rabi coupling leads to a reduced velocity selectivity.
\end{references}
\end{document}